# Tracking Solar Active Region Outflow Plasma from its Source to the near-Earth Environment

J.L. Culhane[1] • D.H. Brooks[2] • L. van Driel-Gesztelyi[1,3,4] • P. Démoulin[3] • D. Baker[1] • M.L. DeRosa[5] • C.H. Mandrini[6,7] • L. Zhao[8] • T.H. Zurbuchen[8]

[1]UCL-Mullard Space Science Laboratory, Holmbury St Mary, UK
email: j.culhane@ucl.ac.uk
[2]George Mason University, Fairfax, VA, USA,
[3]Observatoire de Paris, LESIA, UMR 8109 (CNRS), Paris, France,
[4]Konkoly Observatory of the Hungarian Academy of Sciences, Budapest, Hungary,
[5]Lockheed Martin Solar and Astrophysics Laboratory, Palo Alto, CA, USA,
[6]Instituto de Astronomia y Fisica del Espacio (IAFE), CONICET-UBA, Buenos Aires, Argentina,
[7]Facultad de Ciencias Exactas y Naturales (FCEN), UBA, Buenos Aires, Argentina,
[8]Dept. of Atmospheric, Oceanic and Earth Sciences, Univ. of Michigan, Ann Arbor, MI, USA

**Abstract**

Seeking to establish whether active region upflow material contributes to the slow solar wind, we examine in detail the plasma upflows from Active Region (AR) 10978, which crossed the Sun's disc in the interval 8 to 16 December 2007 during Carrington rotation (CR) 2064. In previous work, using data from the *Hinode/EUV Imaging Spectrometer*, upflow velocity evolution was extensively studied as the region crossed the disc while a linear force free-field magnetic extrapolation was used to confirm aspects of the velocity evolution and to establish the presence of quasi-separatrix layers at the upflow source areas. The plasma properties, temperature, density and first ionisation potential bias [FIP-bias] were measured with the spectrometer during the disc passage of the active region. Global potential field source surface (PFSS) models showed that AR 10978 was completely covered by the closed field of a helmet streamer that is part of the streamer belt. Thus it is not clear how any of the upflowing AR-associated plasma could reach the source surface at 2.5 $R_\odot$ and contribute to the slow solar wind. However a detailed examination of solar-wind *in-situ* data obtained by the *Advanced Composition Explorer* (ACE) spacecraft at the $L_1$ point shows that increases in $O^{7+}/O^{6+}$, $C^{6+}/C^{5+}$ and Fe/O – a FIP-bias proxy – are present before the heliospheric current sheet crossing. These increases, along with an accompanying reduction in proton velocity and an increase in density are characteristic of both AR and slow-solar wind plasma. Finally we describe a two-step reconnection process by which some of the upflowing plasma from the AR could reach the heliosphere.

**Keywords:** Active region upflow, magnetic topology, slow solar wind contribution

# 1. Introduction

Following the launch of *Hinode* (Kosugi *et al.*, 2007) in September, 2006, upflows, persistent for several days, were observed from active regions (Sakao *et al.*, 2007; Harra *et al.*, 2008) with the *Hinode X-ray Telescope* (XRT: Golub *et al.*, 2007) and the *EUV Imaging Spectrometer* (EIS: Culhane *et al.*, 2006) instruments. These upflows have attracted considerable attention since it was estimated that they could supply a significant contribution to the slow solar wind (Sakao *et al.*, 2007). Persistent upflows of plasma from active-region peripheries with low radiance and comparatively strong associated magnetic field of a single polarity have been reported by Doschek *et al.* (2007, 2008), Del Zanna (2008), and Harra *et al.* (2008). Velocities, obtained from blueshifts, range from $\approx 3$ km s$^{-1}$ to $\approx 50$ km s$^{-1}$ and increase with temperature (Del Zanna, 2008). Non-thermal line broadening and the appearance of blue wing asymmetries are observed to correlate with velocity (Del Zanna, 2008; Harra *et al.*, 2008; Bryans, Young, and Doschek, 2011). This could suggest the presence of multiple upflow sites for plasma at different speeds. Warren *et al.* (2011) examined the temperature structure and morphology of the upflow regions. They found that while the upflow plasma has a temperature [$T_e$] in the range 1 MK $\leq T_e \leq$ 2.5 MK, down-flowing plasma at $T_e \approx 0.63$ MK (Si VII) is detected in coronal fan loop structures and has a different morphology from that of the upflowing material. Thus the latter is not associated with the fan loop population.

Since the identification of the peripheral upflow plasma several driving mechanisms have been proposed. These include coronal plasma circulation due to open magnetic field funnels (Marsch *et al.*, 2008), impulsive heating at AR loop footpoints (Harra *et al.*, 2008), chromospheric evaporation following reconnection driven by flux emergence and braiding due to photospheric motions (Del Zanna, 2008), expansion of large-scale reconnecting loops (Harra *et al.*, 2008) and continual AR expansion (Murray *et al.*, 2010). The latter is the only mechanism that does not involve magnetic reconnection. Baker *et al.* (2009) have pointed out that in all of the cases involving active-region peripheral upflows that have been studied to date, the upflows appear at sites where magnetic-field lines of strongly different connectivity are rooted or meet. These sites are called quasi-separatrix layers (QSLs; Démoulin *et al.* 1996), and in the special case involving discontinuity of the field-line connectivity, *e.g.* between closed and open fields, they become separatrices. QSLs are thin 3D volumes (surfaces in the limit of separatrices) and they are preferential locations for magnetic reconnection. This can occur between AR loops and either longer lower-density loops that connect to other parts of the Sun or with open magnetic field lines. In the latter case, if the field lines reach the source surface they can supply active-region material to the solar wind. This does not necessarily occur for all AR peripheral upflow sites. Using a global potential-field source-surface (PFSS) model, van Driel-Gesztelyi *et al.* (2012) demonstrated that from a peripheral coronal upflow site, which was linked to a coronal null, AR plasma could be channelled to the solar wind *via* interchange reconnection with adjacent open field from a coronal hole (CH). These authors found evidence from the *Advanced Composition Explorer* (ACE) *in-situ* data for the presence of active-region material in the slow solar wind that was associated with the passage of the two active regions across the disc in January 2008. Presence of (additional) ejected plasmoid material in the heliospheric plasma sheet (HPS) was suggested by Foullon *et al.* (2011).

Composition in particular is key to understanding the origins of the solar wind (Geiss, Gloeckler, and von Steiger, 1995). The fast wind [$V \geq 600$ km s$^{-1}$] is believed to originate in coronal holes and exhibits a composition characteristic of the photosphere with charge states that indicate temperatures in the 0.8 MK to 1.0 MK range (von Steiger *et el.*, 2000). The origins of the slow wind [$V \approx 400$ km s$^{-1}$] remain controversial. The charge states present indicate temperatures > 1.2 MK. In addition the abundances of elements with first ionisation potential (FIP) < 10 eV, *e.g.* Si, Mg, Fe, are enhanced by a factor two to four compared to those with larger FIP, *e.g.* O, S, while the overall slow-wind composition is more variable than that of the fast wind (Zurbuchen and von Steiger, 2006). In the Sun, enhanced low-FIP element abundance, referred to as the FIP bias, is found in particular in active-region plasma. See Laming (2004) for a discussion of the FIP effect and Baker *et al.* (2013) for a spatially resolved *Hinode*/EIS FIP analysis of an active region. The size of the FIP bias appears to be related to the plasma-confinement time in the closed active-region structures (Feldman and Widing, 2003). Established active regions of age > three to four days, can exhibit FIP bias values in the range three to five, so if such plasma were released to the heliosphere along open field lines it could provide a contribution of appropriate composition to the slow wind. Using EIS to observe the emission lines of low and high-FIP elements in the plasma from the upflow sites, the plasma can be characterised for later comparison with *in-situ* measurements made from spacecraft in the heliosphere, *e.g.* ACE. Feldman *et al.* (2009) have outlined how observations of Si X (low FIP) and S X (high FIP) lines in the EIS spectral range could be used to measure the FIP bias. This approach, which was further developed by Brooks and Warren (2011), is used in the present article.

In the following sections we examine the East (E) and West (W) outflows from AR 10978, a region on the disc in December 2007. In Section 2, following an outline of the EIS upflow velocity measurements, we describe the velocity evolution that takes place as the region crosses the disc. A linear force-free magnetic-field model (LFFF) was used to establish the location of QSLs, which coincide with the photospheric footprints of the upflow regions. Measurement of the upflow plasma parameters, with reference in particular to the FIP-bias, is then described and the results displayed. The global magnetic field configuration is presented which shows the relationship of AR 10978 to the heliospheric current sheet (HCS). A detailed global PFSS model was then constructed to allow a search for open field paths by which the upflow plasma could reach the heliosphere. *In-situ* plasma parameters and magnetic field data obtained by the ACE spacecraft at L$_1$ were investigated and the data back-mapped to the Sun. The discussion in Section 3 relates the arrival of active-region plasma at ACE to the presence of a high altitude magnetic null located North of AR 10978. Conclusions are given in Section 4.

## 2. Active Region Upflows from AR 10978

The *Hinode/*EIS instrument observes in two wavelength bands: 171 – 212 Å and 245 – 291 Å. It is a raster-scanning instrument with a maximum field of view (FoV) of 600 arcsec in dispersion by 512 arcsec along the slit. It has 1 arcsec pixels and a spectral resolution (1σ) of 22.3 mÅ. EIS is described in detail by Culhane *et al.* (2007). A series of raster scans of AR 10978 was carried out in the period between 6 and 19 December 2007. Three examples are given in Figure 1. Here the *Hinode/*XRT full-disc solar images were obtained using Ti/Poly filters. They show the passage of AR 10978 across the solar disc and also indicate the presence of two coronal holes, of opposite magnetic polarity, on the E- and W-sides of the active region. The coronal holes are clearly separated from the E- and W-side active region boundaries. The three corresponding velocity maps, also shown in the figure, were obtained from EIS observations of Fe XII/195.12 Å ($T_e \approx 1.4$ MK) emission line profiles. Exposure times were 40 seconds and the duration of each raster was 5.1 hours. Details of the line profile fitting and resulting velocity measurements are given by Démoulin *et al.* (2013), who undertook a comprehensive analysis of the AR velocities during its disc transit. Their work was focussed on the upflow velocities – shown in blue in the bottom row of Figure 1.

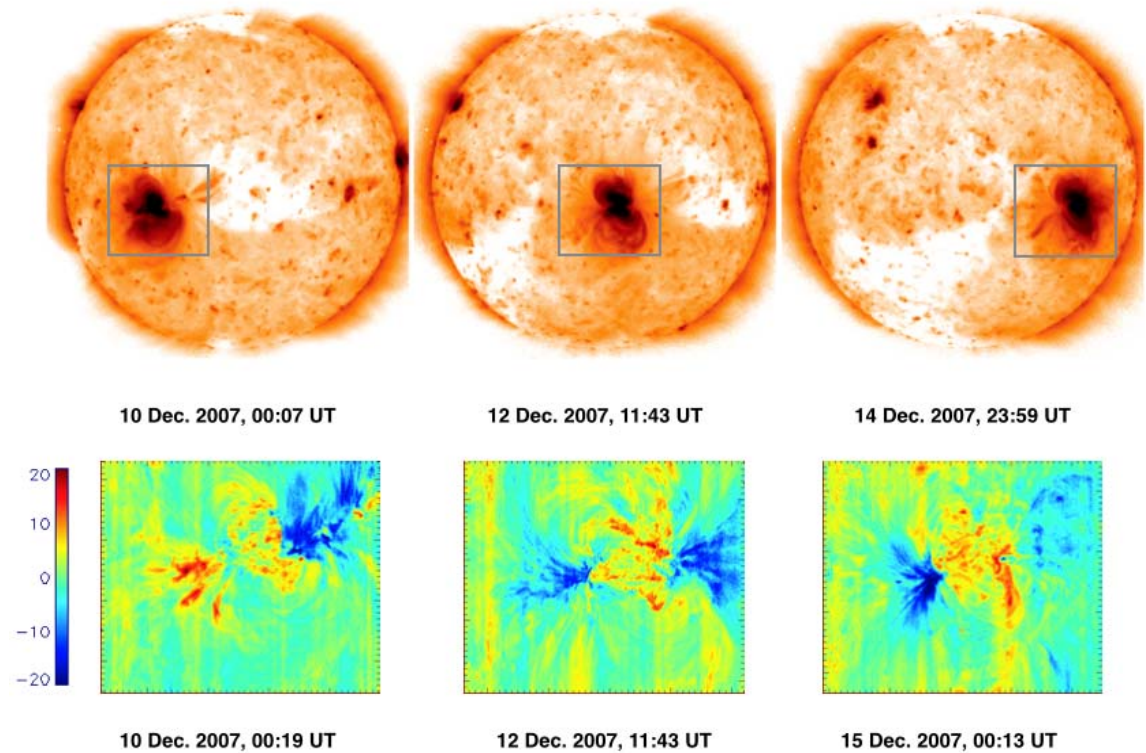

**Figure 1**. AR 10978 images and velocity maps for the interval 10 – 15 December, 2007. Top row: *Hinode*/XRT Ti/Poly X-ray images for 10, 12 and 14 December, 2007 with inverse contrast. Bottom row: Corresponding *Hinode*/EIS velocity maps obtained from Fe XII/195.12 Å emission line profiles. Velocity scale (bottom left) is in km s$^{-1}$. EIS raster size is 460″ by 384″ in the horizontal (*x*) and vertical (*y*) directions respectively.

## 2.1. Upflow Velocity Evolution

As AR 10978 crossed the disc from East to West due to solar rotation, the upflow velocities evolved in a systematic manner. Thus while both eastern and western upflow regions were clearly visible at or near disc centre, the eastern upflow velocities were hardly detectable when the AR was close to the east limb while the western upflow was clearly in evidence. The situation was reversed as the region approached the west limb with the eastern flow becoming dominant. A detailed analysis by Démoulin *et al.* (2013) of the apparent variation of the velocities with time over the half solar-rotation interval clearly showed that this variation can be explained as a projection effect as the AR rotates across the disc. They deduced that the upflows on both eastern and western sides are steady and tilted away from the vertical to the East and West respectively. The situation is summarized in the schematic diagram given in Figure 14 of their article. The upflow velocities were deduced principally from Doppler-velocity shifts in the Fe xii/195.12 Å line profiles.

In order to check the consistency of the inclination results and to investigate the upflow driving mechanism, Démoulin *et al.* (2013) carried out a linear-force-free magnetic-field (LFFF) extrapolation (see Démoulin *et al.*, 1997). Since AR 10978 has a magnetic configuration that is close to potential, the result can be regarded as being reasonably representative. The results of their LFFF extrapolation model are summarised in Figure 7a, b of their article where the photospheric traces of QSLs are also indicated. The upflow directions deduced from the computed field-line inclinations are broadly consistent with those derived from **their** previously described velocity evolution measurements. The upflows originate mainly from the external sides of the East and West QSLs remote from the AR. The origin of persistent active region upflows from QSLs has been previously reported (Baker *et al.*, 2009; van Driel-Gesztelyi et al., 2012). These upflows are thought to be due to the reconnection of comparatively dense active-region loops with surrounding lower-density loops (Bradshaw, Aulanier, and Del Zanna, 2011). Previous observations have shown the upflows to be persistent on a timescale of days where the required ongoing reconnection is driven by AR growth and dispersion. This is consistent with the present observations of AR 10978 where the upflows show short-term variability but with a strong stationary component indicating a long-lived driving mechanism. It was also observed that the upflows from the east side of AR 10978 were comparatively steady during the AR's disc passage while the upflows on the west side showed variations due to the emergence of new magnetic flux in the period following 12 December 2007.

While the LFFF magnetic-field configuration is valid in the neighbourhood of the AR, it is important to emphasise that the existence of AR-related upflows *does not* imply that the AR plasma leaves the Sun and reaches the heliosphere. This issue will be addressed later in the present article and will be examined in detail in a separate article (Mandrini *et al.*, 2014).

## 2.2. Properties of the Upflowing Plasma from AR 10978

Knowledge of plasma parameter values [$n_e$, $T_e$, and composition] is important in tracking the location and flow of plasma in the solar atmosphere. The above parameters were measured **by** Brooks and Warren (2011) for both of the upflow regions associated with AR 10978 during the period 9 to15 December 2007. The emission-line intensity data used in the study were obtained and processed as described in their article using the *Hinode*/EIS instrument.

Density [$n_e$] in the upflow regions was measured using the diagnostic intensity ratio [$I_{Fe\ XIII\ (202.044\ Å)}/I_{Fe\ XIII\ (203.826\ Å)}$] and is shown in Figure 2b for the 9 to 15 December interval. The ratio of emission-line intensities from low-FIP and high-FIP ions can be used to identify coronal plasma and in particular to allow estimates of FIP bias or $f_{FIP} = A_{SA}/A_{ph}$: the ratio of solar atmosphere to photospheric element abundance. Brooks and Warren (2011) showed that the ratio $I_{Si\ X\ (258.375\ Å)}/I_{S\ X\ (264.223\ Å)}$ is constant to within $\approx$ 40% for the log ($T_e$/K) = 5.7 – 6.2 range but deviates significantly at higher temperatures. Hence for reliable estimates of FIP bias a differential emission measure (DEM) analysis based on a range of line intensities, *e.g.* lines from ions Fe VII to Fe XVI, is essential so that proper allowance can be made for the contribution to line emissivities **for** a range of temperatures. In addition, $I_{Si\ x\ (258.375\ Å)}/I_{S\ x\ (264.223\ Å)}$ is sensitive to electron density – with a factor 2.3 variation for the log ($n_e$/cm$^{-3}$) = 8 – 10 range. Hence the electron density in the upflow regions must be measured using the diagnostic ratio mentioned above. Details of the DEM analysis methods used and of the resulting measurement of FIP bias are given by Brooks and Warren (2011). In the present article, we exploit their analysis in a different way by characterizing the temporal evolution rather than spatial variation of FIP bias values based on the Si X to S X line intensity ratio. This evolution is shown in Figure 2c for the 9 to 15 December interval, while the $T_e$ values that characterise the peaks of the DEM distributions are plotted in Figure 2a for the same time interval. In Figure 2, the plasma parameter values for the East (red) and West (green) upflow regions are indicated separately.

FIP-bias values estimated for the AR upflows (Figure 2c), are in the range $3.0 \leq f_{FIP} \leq 4.0$ and show no change in the 11 to 13 December interval while the region is close to disc centre. Simple estimates of the errors (about 25% – 30%) are indicated for the FIP-bias values based on the relative error in predicted to observed intensities for the S X line intensity only. The values are consistent with plasma of coronal origin but the errors do not allow any clear difference between E- and W-upflows to be detected even though on-going flux emergence to the West of AR 10978 is probably bringing lower FIP-bias plasma from the photosphere and a weak tendency to have lower FIP on that side is suggested by the absolute measurements.

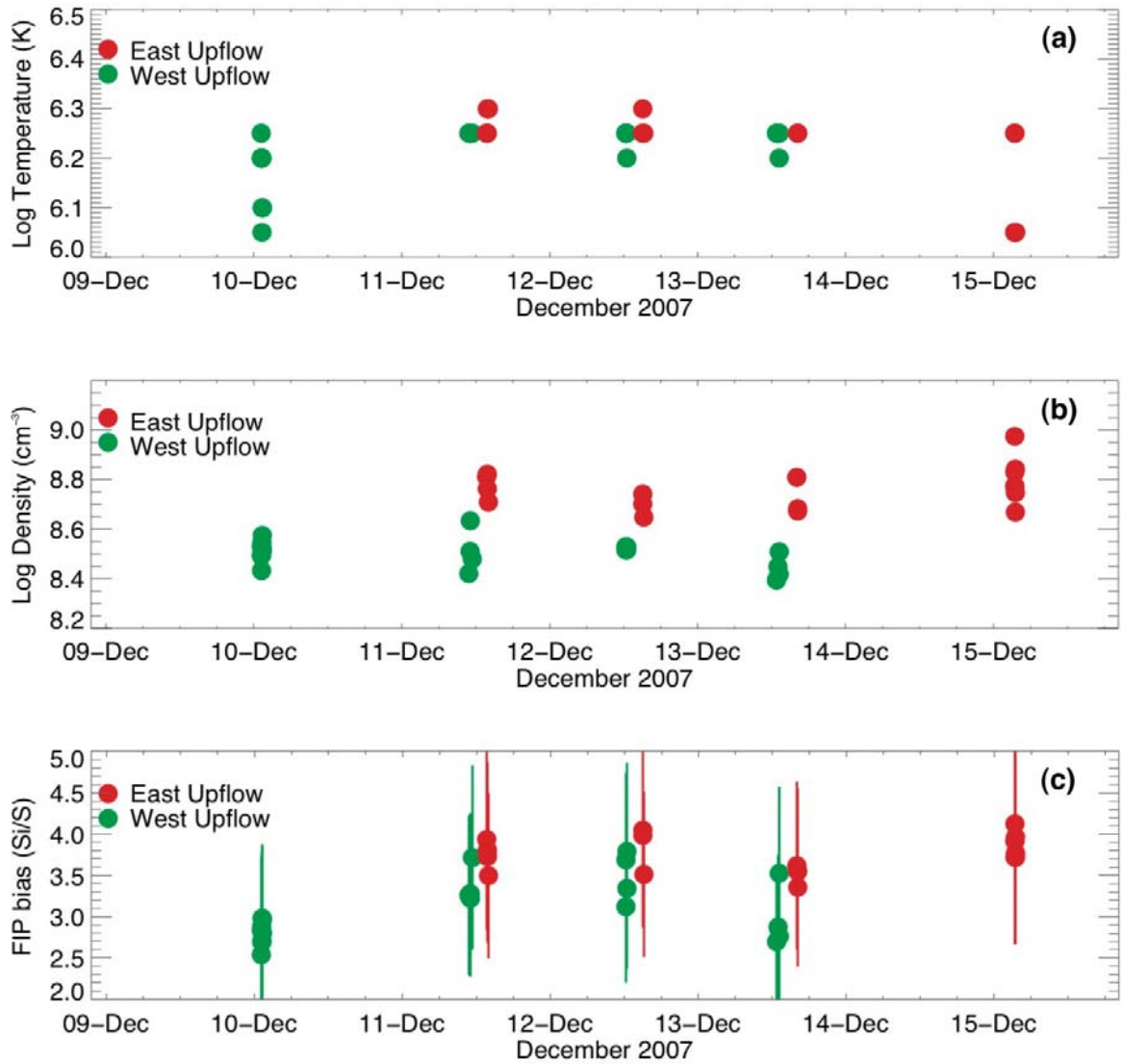

**Figure 2.** Upflowing plasma properties for east and west upflows. Panels a, b, and c correspond to $T_e$, $n_e$, and FIP bias respectively. Parameter values obtained from appropriate line ratios (b) and DEM analysis (a), (c). (See Brooks and Warren (2011), for details of the FIP bias determination).

## 2.3. The Global Magnetic Field Configuration for AR 10978

The time interval 30 November to 27 December 2007 covered by CR 2064 includes the disc transit of AR 10978. A Carrington map for this rotation, constructed with meridional strips from the *EUV Imager* (EUVI) images, on board the *Solar Terrestrial Relations Observatory-B* (ST-B) is shown in Figure 3. AR 10978 is indicated in the figure and is seen to be located between two nearby coronal holes of opposite magnetic polarity. An overall view of the global magnetic context for the region is given by the relevant *Global Oscillation Network Group*, National Solar Observatory (NSO/GONG), PFSS model (Figure 4). In this model the source surface is at 2.5 $R_\odot$. The areas in red and green show the negative (-ve) and positive (+ve) polarities of the open-field regions. The grey areas show the regions of mainly closed magnetic field. The essential line-of-sight magnetogram features of AR 10978 are also shown. The plasma-upflow sites associated with the active region are indicated by red (eastern) and yellow (western) arrows. During CR 2064 the inward projection of the HCS, shown in blue in the figure, clearly separates the east and west upflow regions and in fact essentially forms the magnetic inversion line for AR 10978. The red and green open-field regions include the two coronal holes that are visible in the ST-B EUVI images from the Carrington plot in Figure 3.

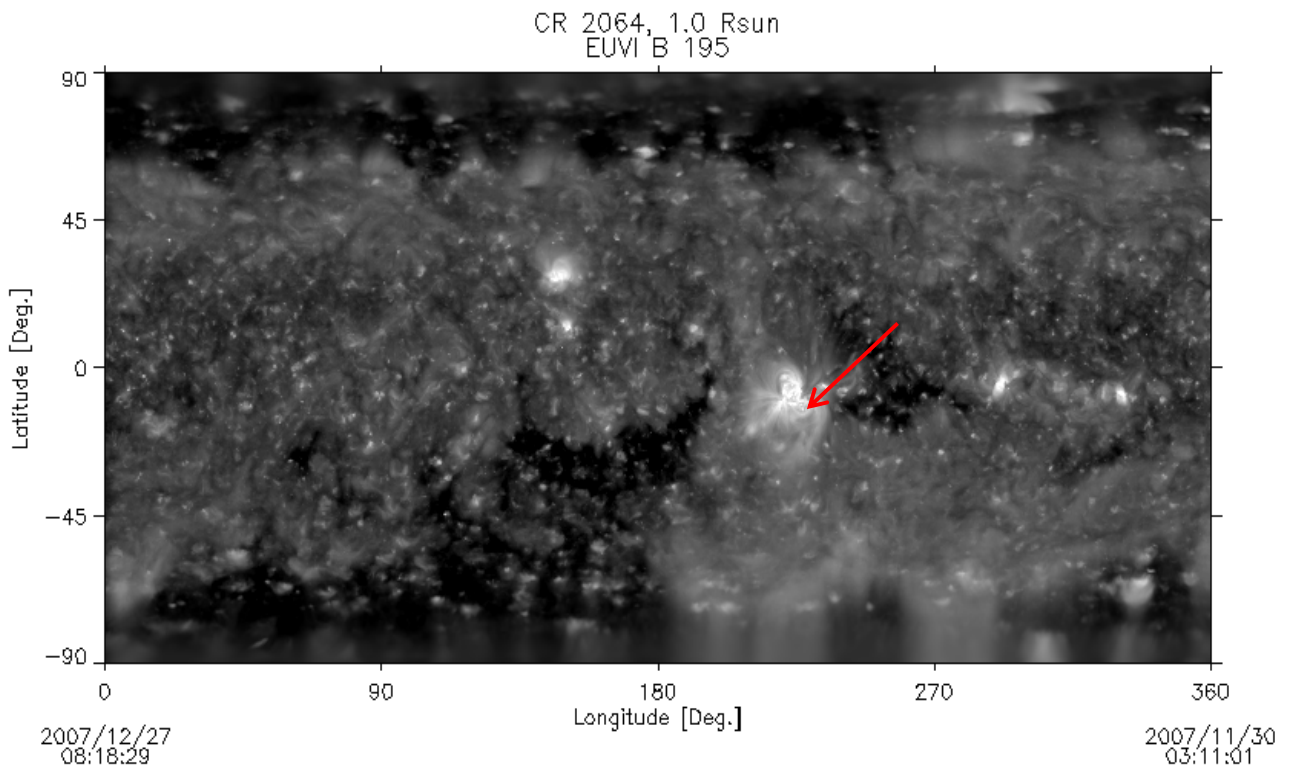

**Figure 3.** Carrington Rotation (CR) 2064 shown with ST-B EUVI images. AR 10978 is indicated with a red arrow. Time increases from right to left.

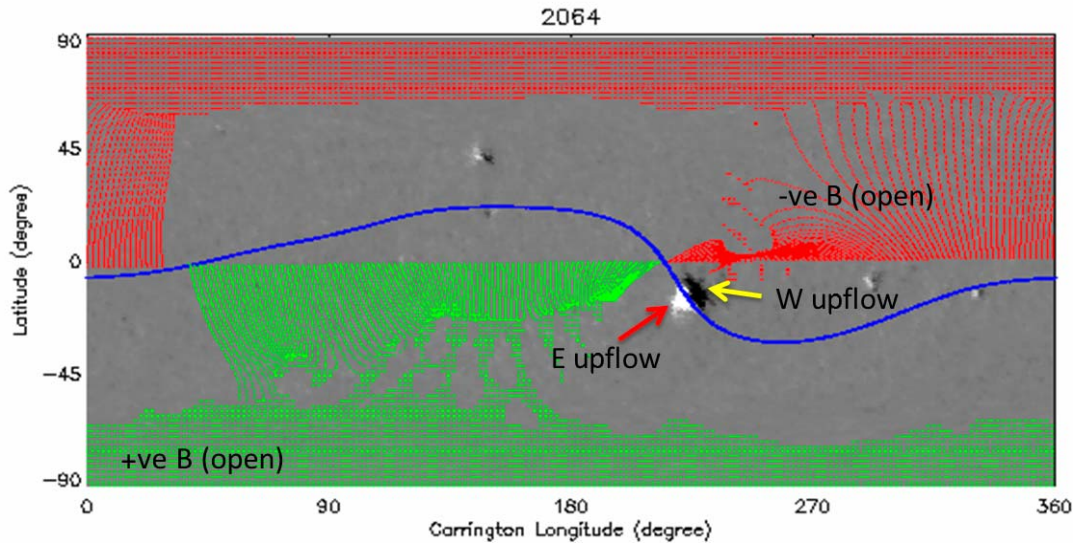

**Figure 4.** NSO/GONG PFSS model with the source surface at 2.5 $R_\odot$. Red(Green) areas show negative(positive) polarity of the open-field regions. The solid-blue line shows the field inversion line at 2.5 $R_\odot$. Grey area indicates mainly closed field regions and includes the line-of-sight (LoS) magnetogram features for AR 10978.

A more detailed study of global magnetic topology was undertaken in an attempt to clarify the ultimate destination of the AR 10978 upflows and in particular to help establish whether or not any of the upflowing plasma can escape into the heliosphere and contribute to the solar wind. The global PFSS model shown in Figure 4 was prepared using synoptic magnetic-field data. A more specific model was constructed for the period around 12 December 2007, when the AR was at or near central meridian. Here the surface magnetic-field measurements are sampled from an evolving flux model (Schrijver and Title, 2001) using the method for inserting the field data into the PFSS model described by Schrijver and DeRosa (2003). The large-scale topological features of this model, *e.g.* separatrix surfaces and null points, are calculated using the methods of Haynes and Parnell (2010) as adapted for spherical geometries and are shown in Figure 5. The null points are indicated by small red dots and the separatrix surfaces are indicated by various semi-transparent surfaces, with each surface having a different color. In particular, the yellow separatrix surface covering the AR shown in Figure 5a is the helmet surface that separates open from closed flux in the model. The dark-blue line at the apex of the helmet surface represents the line of null points at the upper boundary of the PFSS model and is coincident with the contour separating positive and negative open polarities on the source surface at 2.5 $R_\odot$. This line of null points represents the base of the heliospheric current sheet that, on the Sun, is presumed to be surrounded by the helmet streamer structures observed in coronagraph imagery. In Figure 5a, on both sides of the helmet surface, South-East and North-West of the AR, open-field lines occupy the volume between the yellow helmet surface and the neighboring separatrix surfaces. The photospheric footprints of these open-flux regions are the grey areas in Figure 5a where the greyscale magnetic map is unobscured. The photospheric open-flux regions in the PFSS model are related to the two CHs shown in the top panel of Figure 1.

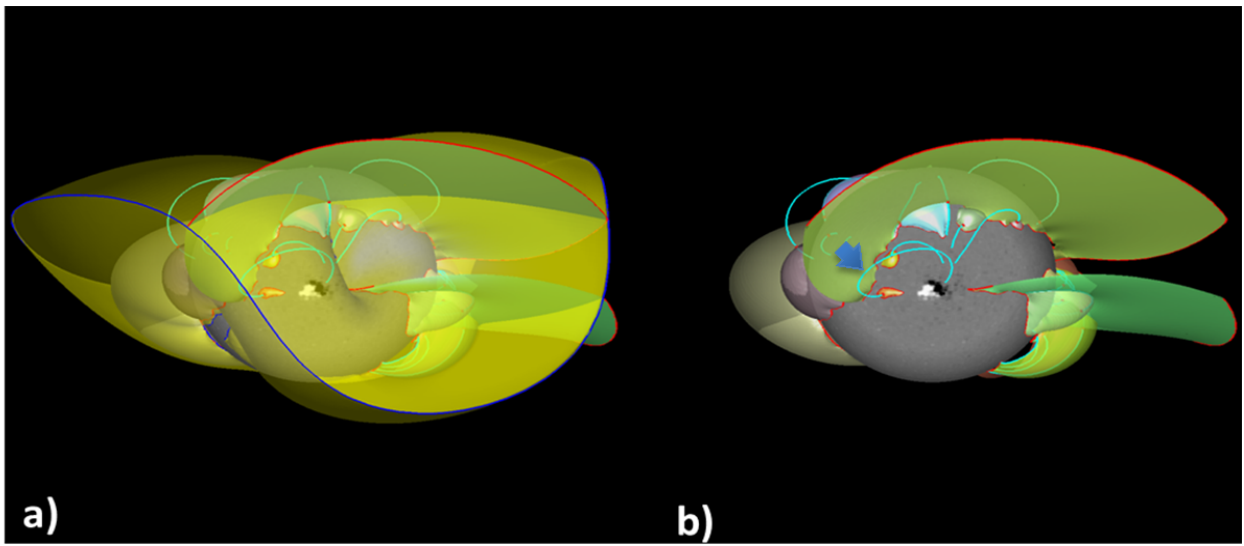

**Figure 5.** PFSS model showing large-scale topological structures on 12 December 2007 around AR 10978. Panel a includes the semi-transparent yellow helmet streamer separatrix surface while (b) does not so as to better show the magnetic configuration around AR 10978. The blue line in (a) shows the HCS (*c.f.* Figure 4). The AR is entirely covered by the streamer with no topological link between plasma upflows and open field. The spine field line (cyan; indicated by the blue arrow in panel b) of a low-altitude null-point (red dot) to the East of the AR, which is linked to the western upflow region, indicates that the upflowing plasma is channelled towards quiet-Sun regions along long loops.

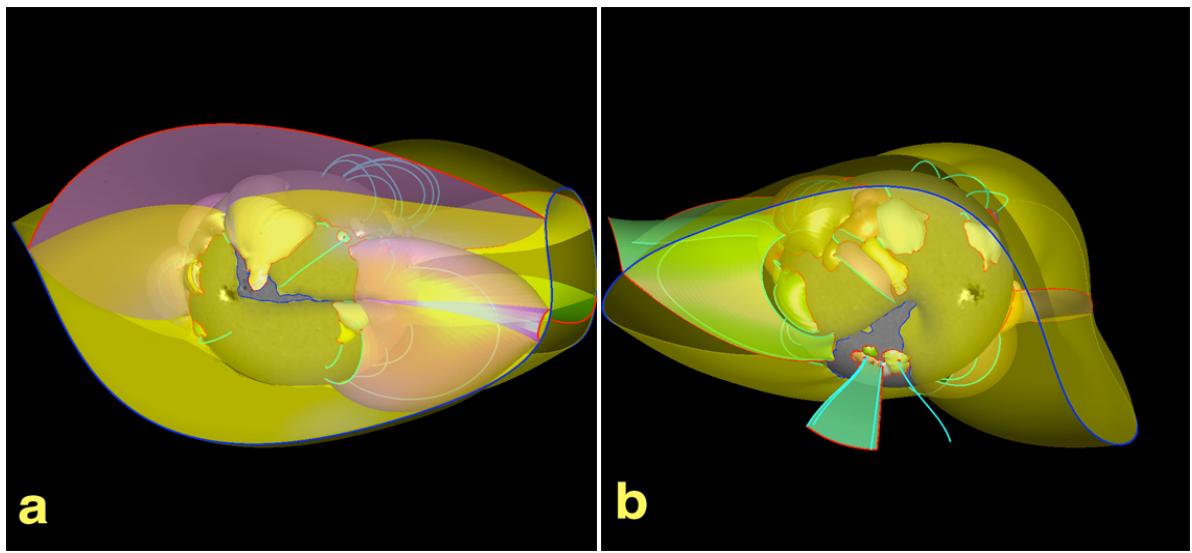

**Figure 6.** PFSS model for 9 and 15 December 2007 (panels a and b respectively) to provide an additional global perspective, showing two open-field (CH) areas flanking AR 10978. As in Figure 5, the main, yellow, surface is the PFSS helmet, *i.e.* the envelope of the domain in the solar corona within which all field is closed. At its cusp on the source surface, indicated by a blue magnetic-null line, the radial-field component vanishes, so that this cusp forms the basis of the HCS.

They are also apparent in the EUVI images shown in Figure 3. A further PFSS model for 9 December 2007, when the AR is East of central meridian, is shown in Figure 6a and on 15 December 2007, when the AR is West of central meridian in an effort to better display the complex 3D structures involved. The PFSS models show the progressive evolution and comparative stability of the global structures during the interval for which AR 10978 is within $60^0$ of central meridian.

From Figures 5a and 6a we see that the AR is fully covered by the streamer. In Figure 5b, the topology is shown without the semi-transparent yellow surfaces for clarity. The light-blue (cyan) spine field lines do not enter the open-field domain but close under the s*treamer surface.* Note that this configuration is quite different from the quadrupolar large-scale topology found one rotation later by van Driel-Gesztelyi *et al.* (2012), where reconnection at a high-altitude null point directs part of the upflowing plasma into the solar wind. This null point was due to a new emerging AR on the east side of AR 10978. In the present simpler bipolar configuration, with the streamers around the HCS completely enclosing the AR, there are no high-altitude null points**.** Thus it is not clear how material from either the eastern or the western AR upflows could enter the heliosphere. However in Section 2.4 we will show that material of AR composition does in fact appear in the *in-situ* solar-wind data as observed by ACE. The plots of the ACE plasma data with respect to the registration of the HCS crossing by the ACE magnetometer are consistent with the solar-origin site of the plasma stream being just to the West of the HCS.

**2.4 ACE *in-situ* Observations and their Possible Relation to AR Outflows**

To search for evidence of active-region plasma in the solar wind, we have examined the ACE *in-situ* data around the time that the magnetometer (*Magnetic Field Experiment* (MAG): Smith *et al.,* 1998) registered a change in field direction due to encountering the HCS. *In-situ* observations obtained by ACE for the plasma flow passing the spacecraft include the ratios Fe/O, $O^{7+}/O^{6+}$, $C^{6+}/C^{5+}$, and $He^{2+}/H^+$ measured with the *Solar Wind Ion Composition Spectrometer* (SWICS: Gloeckler *et al.*, 1998) along with proton velocity and density from the *Solar Wind Electron Proton Alpha Monitor* (SWEPAM: McComas *et al.,* 1998). These quantities have been back-mapped to the source surface 2.5 $R_\odot$ above the photosphere and are plotted on a Carrington display scale for CR 2064 in Figure 7. In order to facilitate the comparison with the solar structures, the ACE data are displayed along with the EUVI image data from Figure 3. Time is running from right to left. The HCS location – shown by the vertical red line in Figure 7 – is established from the magnetometer data. This provides a useful reference for linking the ACE observations to the solar source surface. Mapping the *in-situ* observations of solar wind to the Sun was done in two steps: i) a radial, ballistic mapping from the spacecraft to the 2.5 $R_\odot$ source surface, at which the magnetic field is assumed radial by the PFSS model; ii) a PFSS extrapolation to continue the solar-wind map from source surface to the photosphere. The two stage back-mapping process is illustrated schematically in Figure 8.

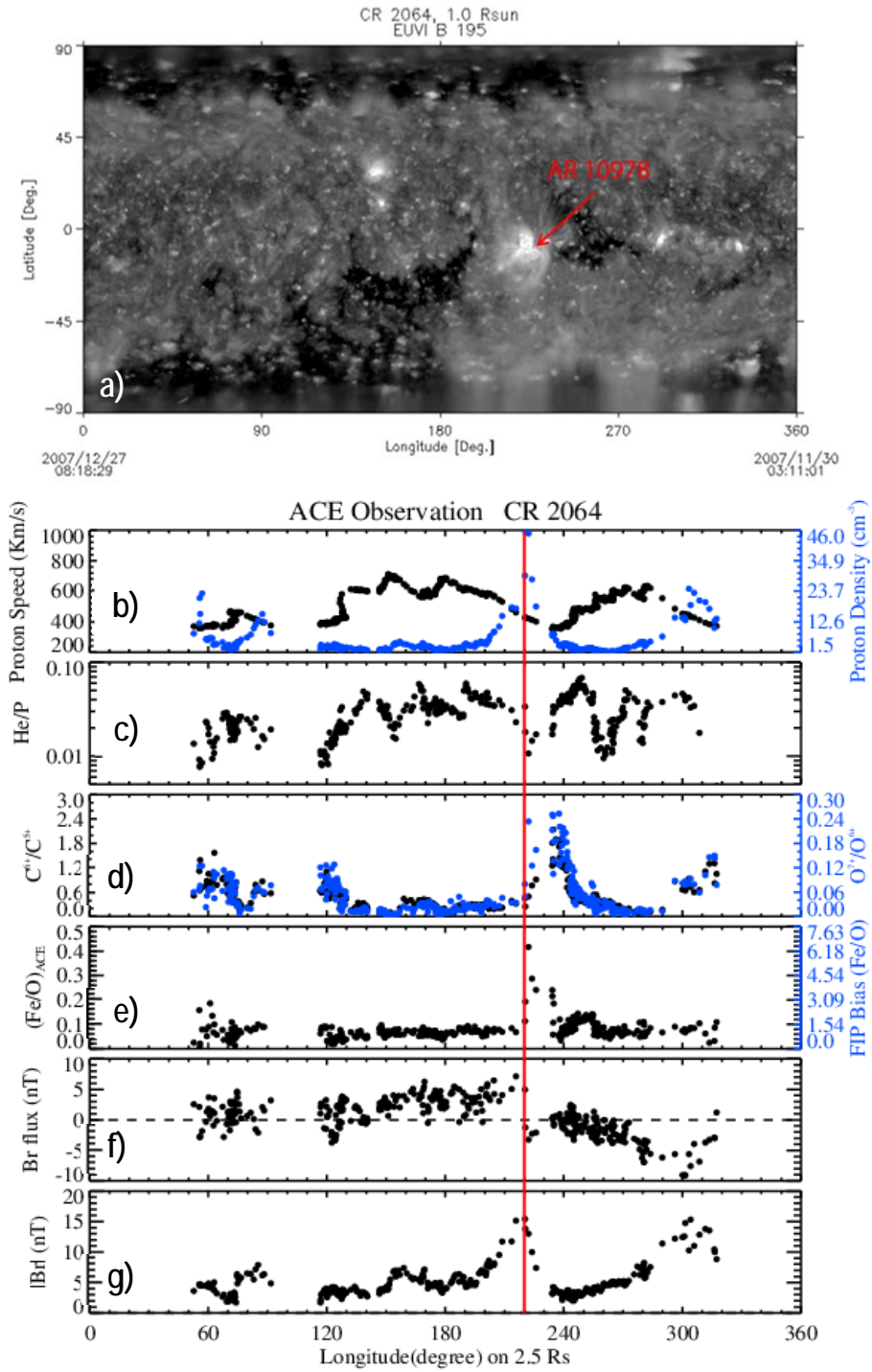

**Figure 7**. ST-B EUVI 195 Å synoptic map for CR 2064 with AR 10978 indicated (a). ACE data back-mapped to 2.5 R$_\odot$ include**:** b) proton speed and density**,** (c) He/p, (d) O$^{+7}$/O$^{+6}$ and C$^{+6}$/C$^{+5}$ ratios, **e**) Fe/O and FIP bias, (f) radial magnetic field [$B_r$] and (g) absolute magnetic field [|B|]. The HCS (indicated by a red line) is between two fast-wind streams from the CHs. Slow-wind signatures – increases in the C and O ion ratios and FIP bias (Fe/O) – are present from the west side of the HCS.

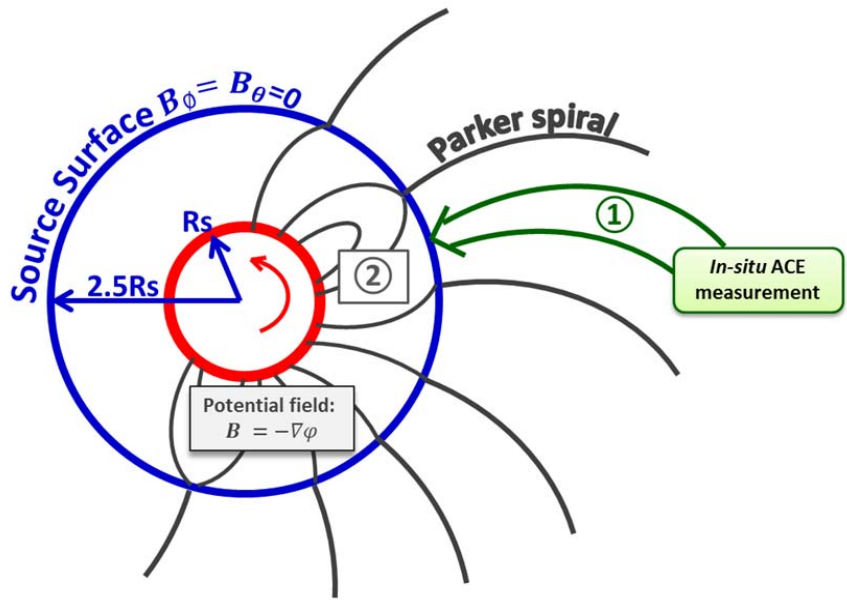

**Figure 8.** Sketch of the two-step mapping technique. Outside of the source surface (2.5 $R_\odot$), the magnetic field lines follow the Parker spiral. The *in-situ* measurements are ballistically mapped back to the source surface (mapping ①). Inside the source surface, we use the PFSS model to track the field to the photosphere (mapping ②).

In the first step, we assume that the solar-wind plasma propagates radially from the source surface to the spacecraft at a constant speed equal to the observed proton speed (Neugebauer *et al.*, 2002). The longitude of the solar-wind source region on the source surface should equal the mapped longitude added to the number of degrees that the Sun rotates during the solar-wind propagation time from the source surface to ACE at the $L_1$ point. The latitude of the solar-wind source region on the 2.5 $R_\odot$ source surface is the same as the heliographic latitude of the spacecraft. This ballistic mapping method has been used and fully evaluated by Neugebauer *et al.* (2002; 2004) with *Ulysses* observations. The advantages and possible uncertainties of this method have also been discussed by these authors. The approach has known shortcomings, especially near interfaces of fast and slow solar-wind streams (Riley *et al.*, 1999). However, as discussed by Neugebauer (2002), the expected errors in source position are approximately ± 10º. This level of uncertainty is supported by the work of Nitta and De Rosa (2008) who considered the location of open-field lines in relation to Type-III radio bursts. Previous studies that used this technique have been described by Zhao, Zurbuchen, and Fisk, 2009; Zhao, Gibson, and Fisk, 2013a,b and by Zhao and Fisk (2011).

In the second step, by using the PFSS software and the Michelson Doppler Imager (MDI: Scherrer *et al.*, 1995) data (Schrijver and DeRosa, 2003), we can continue mapping the solar wind and magnetic field down to the photosphere following PFSS coronal-field lines. There are a number of model-dependent factors that are known to affect the distribution of open *vs.* closed magnetic flux in photospheric-field extrapolations (*e.g.* Poduval and Zhao 2004, Lee *et al.*, 2011). Temporal evolution may also affect the accuracy of the models, which assume time-invariance of the photospheric boundary. Perhaps most important, however, may be the issue of polar magnetic fields. Due to the line-of-sight projection, these are poorly resolved, and are generally reconstructed by incorporating some sort of correction (Arge and Pizzo, 2000). Moreover, the PFSS model makes the undoubtedly incorrect assumption that coronal currents are insignificant. However, it still provides a good starting point for identifying solar-wind source regions. For example, a validation of this two-step mapping technique has been given by Gibson *et al.* (2011) and by Zhao, Gibson, and Fisk, 2013b, where the observed $B_r$ component is first ballistically mapped back to 2.5 $R_\odot$ surface and then tracked back by PFSS model to the solar surface. The longitudes of the polarity-switching points of $B_r$ and the HCS crossings at the ACE's trajectory agree reasonably well.

The data displayed in Figure 7 have been back-mapped to the source surface as described above for the first step. The coronal holes SE and NW of the active region should be sources of fast solar wind. This is indeed reflected in the values of proton speed and density shown in Figure 7b. Plasma speed falls to slow-wind values on either side of the HCS crossing which is detected as a change in field direction by the ACE magnetometer (Figure 7f). Absolute magnetic field is also plotted (Figure 7g). Significant increases in plasma ionisation state, indicated by the increases in the $O^{+7}/O^{+6}$ and $C^{+6}/C^{+5}$ ratios (Figure 7d), are also characteristic of the slow solar wind and suggest "freeze-in" temperature values consistent with an active-region origin for the solar source material. In addition the Fe/O ratio (Figure 7e) is plotted on the ACE and FIP-bias scales for comparison with the FIP-bias parameter displayed in Figure 2c for the AR upflow plasma. In the period when AR 10978 is passing central meridian, the FIP-bias for the eastern upflow in particular (see Figure 2c) is in the range 3.5 to 4.0 – characteristic of closed active region structures, compared to the photospheric value of 1.0. The Fe/O ratio measured by ACE (Figure 9e) likewise increases by a factor of three to four just before the HCS crossing signature is detected. In the same period, the He/p ratio (Figure 7c) falls, which is consistent with the passage of ACE across the streamer belt (Borrini *et al.*, 1981). The increase in Fe/O registered by ACE strongly suggests that upflowing plasma from AR 10978 reached the near-Earth environment. Furthermore the plasma density is a factor of three larger before compared to after the HCS crossing, in contrast to the magnetic-field strength (Figure 7g), indicating that compression is not the only origin of the higher plasma density before the HCS crossing. Thus the slow solar wind plasma sampled by ACE before the HCS crossing appears to have a significant contribution from upflowing active region plasma. The plasma stream also shows the reduction in He abundance associated with streamer belt crossings.

The proton speed measured by ACE is shown in Figure 9 by a set of coloured traces, where colour indicates the speed value. Here the back-mapping has been extended to the photosphere as described above for the second step. Speeds characteristic of the fast solar wind, shown by light-green, yellow, and red traces, originate in the coronal holes that lies to the East and West of the active region. Light and dark blue traces indicate that slow solar wind originates from the site of AR 10978. In the figure, the HCS projection on the EUVI image is illustrated by the blue-dotted curve while the black trace at $0^0$ latitude shows the projection of the ACE trajectory. The red and blue-dotted line at the top of the figure displays the change in magnetic-field direction as the ACE footprint crosses the HCS.

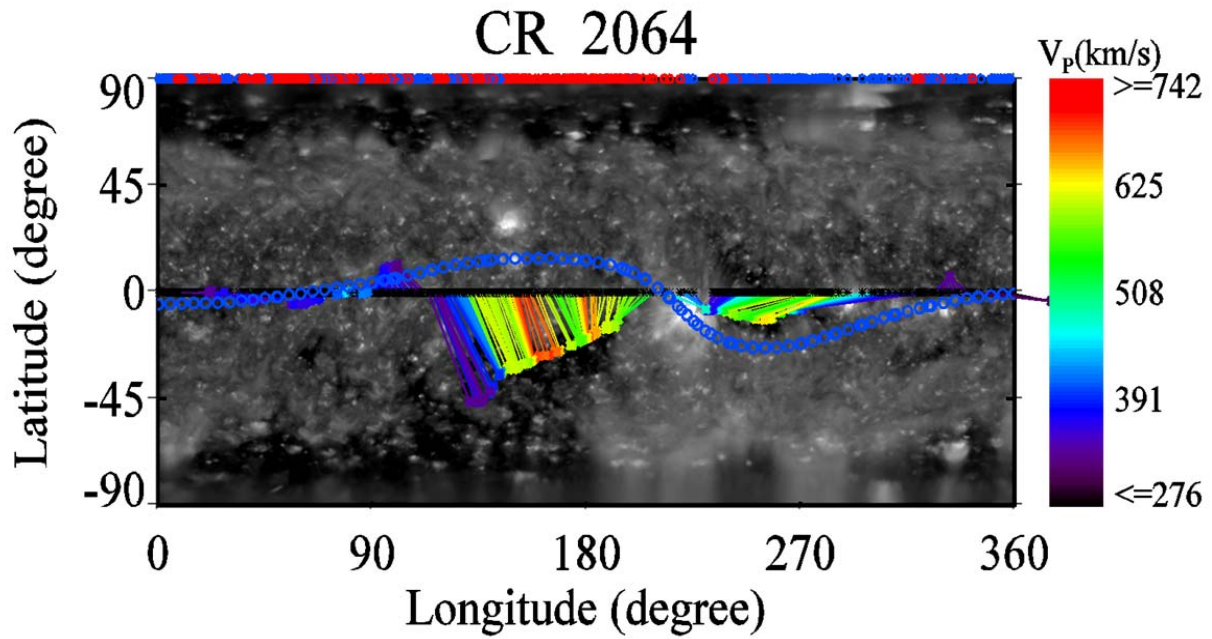

**Figure 9.** ACE data displayed on a STEREO ST-B EUVI 195 synoptic map for CR 2064. Black trace shows ACE trajectory. Circles at top of image show the HCS. $B_r$ sectors at ACE; red/+ve and blue/-ve. Proton speed from ACE is shown by coloured traces - red is fast wind, blue is slow. The traces link the ACE trajectory to the deduced solar source using the back-mapping technique described in the text.

An alternative approach to linking remotely sensed solar observations from the *UV Coronagraph Spectrometer* (UVCS: Kohl *et al.*, 1995, 1997), onboard the *Solar and Heliospheric Observatory* (SOHO), with near-Earth *in-situ* data obtained by ACE has been pioneered by Ko et al. (2001, 2006) who combined ballistic back-mapping to 30 $R_\odot$ with a 3D MHD simulation of the corona below this level. This indicated that during CR 1955 (October, 1999), a contribution to the slow solar wind came from a site located between an equatorial coronal hole and a complex active region. We have applied our two-stage back-mapping technique, as used to obtain the result in Figure 9 above, to the ACE data from CR 1955 and obtained essentially the same outcome as that shown by Ko et al., (2006): their Figure 8. The corona above this site was sampled in some detail by the UVCS line-of-sight passing through the relevant plasma when the region was at the west limb, some seven days after the region crossed disc centre and released the plasma later detected by ACE. By occulting the disc, UVCS allows the detailed sampling of plasma properties above the AR/CH site with the possible disadvantage that the sampled plasma is observed about seven days after the material emitted at disc centre that was observed at ACE**.**

In CR 2064, the presence of active-region upflow plasma in the slow solar-wind stream detected by ACE, as indicated in particular by the factor of four increase in the Fe/O (FIP bias) ratio, is difficult to understand in light of the global magnetic topology indicated by the PFSS extrapolations shown in Figures 5 and 6. As was described in Section 2.3, AR 10978, including both E- and W-upflows, is fully contained below the helmet surface. Thus from the PFSS models, there is no obvious path for plasma of active-region composition and origin to reach the heliosphere and be registered at $L_1$ by the ACE instruments.

**3. Discussion – Possible Explanation for the Detection of AR10978 Plasma by ACE**

Following the early *Hinode* observations of AR-associated upflows (Sakao *et al.*, 2007; Harra *et al.*, 2008), the issue of a possible contribution by this plasma to the solar wind was immediately raised. Given that active-region material was involved and that its composition is similar to that of the slow solar wind, it seemed likely that the upflowing plasma would be associated with the latter. Sakao *et al* (2007) suggested that as much as 25 % of slow-wind plasma could be supplied by active regions, but they did not identify the possible paths by which such material could reach the heliosphere. In their study of AR 10978, Brooks and Warren (2011) established that an enhanced FIP bias for the upflowing plasma persisted for the interval 10 – 15 December 2007. This value was in agreement with the daily average values of the Si/S ratio measured by ACE on 13 – 14 December – a delay consistent with the Sun-to-Earth travel time of the slow solar wind. As was pointed out by Brooks and Warren (2011), this is suggestive of a significant contribution by the AR 10978 plasma to the slow wind while the region was close to disc centre. However they did not discuss possible upflow plasma pathways to the heliosphere, and the global magnetic topology illustrated by the model results displayed in Figures 5 and 6 shows no obvious path for the necessary plasma outflow to the solar wind in the neighbourhood of AR 10978. It is possible that a weak component of the western outflow could lie on open-field lines, but even in this case it is unclear what mechanism could provide coronal plasma directly to the solar wind.

Magnetic nulls are the topological structures in the solar corona at which magnetic reconnection is likely to occur. Van Driel-Gesztelyi *et al.* (2012) identified a topology in which open field at a high-altitude null point could carry active-region plasma to the solar wind. Mandrini *et al.* (2014) describe a detailed analysis of the present magnetic topology in which they identify four high-altitude magnetic nulls located at heights $\geq 25$ Mm above the photosphere and within $\pm 20°$ longitude of the active region. Three of these are associated to closed field lines but the most northerly null, at a height of $\approx 108$ Mm above the photosphere, has associated open-field lines. As was described in Section 2.1, the original upflows are most likely due to reconnection at QSLs between high-density active region loops and surrounding lower-density loops that connect to the network. Continuing AR expansion can drive at the QSLs the comparatively steady upflows over a period of several days (Baker *et al.*, 2009).

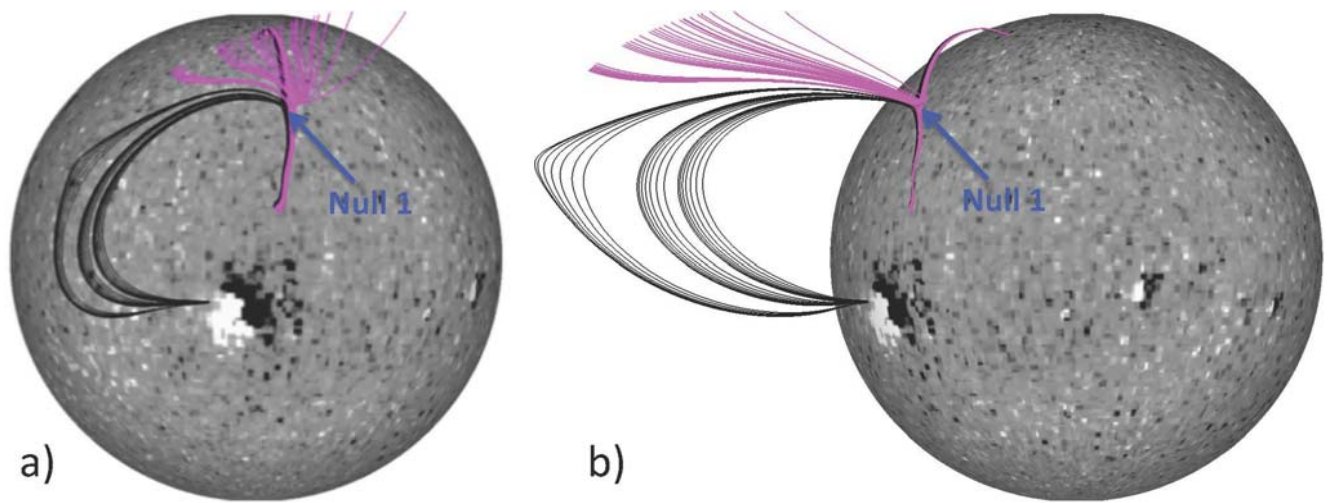

**Figure 10.** The field lines associated with Null 1. Closed-field lines (black) and open-field lines (pink) are indicated. (a) shows AR 10978 at disc centre. (b) a view with the Sun rotated ≈ three days earlier. As open field lines are closer to the separatrix between open and closed field lines, their interplanetary side shifts towards the ecliptic (Adapted from Mandrini *et al.*, 2014)

Mandrini *et al.* (2014) show that, in a first step, the AR expansion forces reconnection at QSLs between the AR closed loops and large-scale network fields at the East of the region. The result of this process will connect the positive field, to the East of the AR, to the network at the North-West of the region. In a second step, additional diffusion of the photospheric AR field forces interchange reconnection at null N1 of the large-scale field lines anchored to the North-West of the AR with the open ones in the neighborhood of the northern negative CH. The magnetic configuration in the surroundings of null N1 is summarised in Figure 10. The two views in Figure 10 indicate that, as interchange reconnection proceeds, the open reconnected field lines will bend towards the Ecliptic. Therefore, the original upflowing plasma from the East side of AR 10978 (Figure 1) is first released in large-scale loops which later reconnect with open field and finally some of the AR plasma is detected *in situ* by ACE before the HCS passage (Figure 7). The ACE detection of the plasma began on 17 December or five days after the AR reached Sun centre. This interval should be around three days for a typical slow-solar wind speed but the longer elapsed time can be explained by the addition of plasma travel time from AR 10978 to Null 1 (Figure 10 a), thus further supporting the suggestion of Mandrini *et al.* (2014).

## 4. Conclusions

Following their discovery by the *Hinode* spacecraft, persistent upflows from active region peripheries have attracted considerable attention not least because of the potential contribution that this plasma could make to the slow solar wind. In the work reported here, we seek to establish whether the upflow plasma associated with AR 10978 could gain access to the heliosphere and thus form a part of the slow solar wind.

We have examined AR 10978, which crossed disc centre on 12 December 2007. Line-of-sight upflow velocities in the range 12 km s$^{-1}$ to 22 km s$^{-1}$ were observed near disc centre while the line profiles can be interpreted as indicating a significant velocity range up to ≈ 100 km s$^{-1}$. Observation of the velocity evolution as the region crossed the disc between 9 and 15 December showed that both eastern and western upflows were inclined to the local vertical at ≈ -50$^0$ and ≈ +20$^0$ respectively. This was independently verified by LFFF magnetic-field extrapolation for AR 10978 and its immediate surroundings. The model also allowed identification of QSLs at the eastern and western edges of the region. Both upflows are rooted in these QSLs consistent with previous active-region upflow observations. They are driven by reconnection at the QSL sites between higher-density AR loops and long lower-density loops that leave the AR neighbourhood. The eastern upflows are steadier than those from the western site consistent with the emergence of new magnetic flux to the West of the region after 12 December.

Although the two principal solar-wind components - slow and fast - are initially characterised by their speeds, it has emerged that plasma composition provides a more reliable and consistent way to differentiate between them. We therefore undertook a detailed study of the plasma properties for the east and west upflow streams. Enhancement of abundance for elements of low first ionisation potential in the solar atmosphere compared to the photosphere – the FIP – bias or $f_{FIP}$ = $A_{SA}/A_{ph}$ - was measured for both of the upflow streams using the ratio of Si x to S x emission line intensities. Reliable estimation of $f_{FIP}$ requires knowledge of plasma temperature and density, so these parameters were also measured using a differential emission measure (DEM) analysis and the density-sensitive line intensity ratio [$I_{\text{Fe XIII} (202.044 \text{ Å})}/I_{\text{Fe XIII} (203.826 \text{ Å})}$]. In the period around 12 December, when AR 10978 was passing central meridian, $f_{FIP}$ was in the range three to four, a value range characteristic of established active regions at about three to four days following emergence and consistent with that found in the slow solar wind. The temperature of the upflow plasma is in the range 1.0 MK to 2 MK as estimated from the DEM analysis. Together with the FIP-bias value this indicates that the upflowing material has the characteristics of AR plasma from closed structures. These features are different from those of the fast solar wind that are typical of photospheric material.

In order to determine whether or not the upflowing material eventually contributes to the slow solar wind, it was necessary to examine the magnetic topology of AR 10978 in the context of the overall solar magnetic configuration at the time of these studies. The region is located between two coronal holes of opposite magnetic polarity. A standard GONG/PFSS model for CR 2064 shows that the inward projection of the HCS essentially bisects the AR and separates the east and west upflow regions. Although nearby the coronal holes do not share common boundaries with AR 10978. More specific PFSS modelling was undertaken in an effort to identify any open-field pathways that could allow the escape of either the eastern or western upflow material to the source surface and beyond. These models showed that the streamer structure below the HCS completely encloses the active region. Any extended magnetic structures leaving the neighbourhood of AR 10978 and connecting to quiet-Sun regions remain closed and are completely covered by the helmet-streamer structure. Thus there are no open field lines that reach the source surface from the immediate neighbourhood of the active region. The strong stationary-upflow component persisted during several days around central meridian passage of the region. In addition several PFSS models showed no obvious change in large- scale topology during the disc passage of AR 10978. Furthermore flare activity in the region was low during this period, and thus no significant modifications to the magnetic field appear to have occurred.

In order to relate the *in-situ* parameters - composition, proton speed and magnetic field - to their possible sites of origin on the solar surface, the ACE data were back-mapped to the Sun in two stages. These were a radial ballistic mapping from $L_1$ to the 2.5 $R_\odot$ source surface, using the *in-situ* measured proton speed and a further PFSS-based extrapolation from the source surface to photosphere. The change in magnetic-field direction when ACE crossed the HCS gives a convenient reference to which the other parameters can be related. For the source-surface backmap, while parameter values characteristic of the fast solar wind are evident before and after the HCS transit, composition indicators ($O^{7+}/O^{6+}$, $C^{6+}/C^{5+}$, and Fe/O – a FIP-bias proxy) increase significantly just west of the HCS by a factor of ≈ four to five, consistent with having entered a slow-wind stream. The He/p ratio falls before the HCS crossing during the same time interval when the Fe/O ratio is enhanced. This is consistent with passage over the streamer belt. Proton speed falls as ACE approaches the HCS and reaches its lowest value to the West of the HCS. Proton density is highest just to the west side of the HCS. Within the ± 10° uncertainty in the back-mapping methods, the ACE *in-situ* observations are consistent with a plasma stream, of active-region composition originating from the neighborhood of AR 10978. Although plasma does leave the region in the east and west upflow streams, the PFSS modelling that was specifically focused on this active region shows no evidence of open magnetic-field structures that could have carried this upflowing plasma to the source surface and beyond to the ACE spacecraft.

Faced with this apparent inconsistency, Mandrini *et al.* (2014) undertook a detailed analysis of the magnetic topology during CR 2064. They sought to identify high altitude magnetic nulls: sites in the corona where magnetic reconnection is likely to occur. Of the four that they identified at heights ≥ 25 Mm, only one located to the North-West of AR 10978, close to the northern coronal hole and at a height of ≈ 108 Mm, had associated open-field lines. They proposed that the AR global expansion forces reconnection between high-density closed loops within the AR and large low-density network loops anchored East of the region. In a second step, the large loops which reach the neighbourhood of the null point can undergo reconnection with open field that originates from the northern coronal hole. At larger solar distances, the reconnected open field bends towards the Ecliptic and can thus deliver plasma that originated in the eastern and western upflows from AR 10978 to the Sun–Earth line where it can be registered by the ACE *in-situ* instruments several days later.

While the work presented indicates that the recently identified AR upflows can contribute to the slow solar wind, it is so far proving to be a valid finding only on a case-by-case basis. In particular for AR 10978 discussed here, it is highly likely that material from the eastern and western upflows is detected at ACE following indirect paths that involved a reconnection process in at least two steps. Van Driel-Gesztelyi *et al.* (2012) studied a pair of active regions with two adjacent coronal holes during the following solar rotation (CR 2065), and found that some of the upflow involved escaped along open field lines associated with a null point. However part of the upflow plasma remained confined. Thus there is a need for continuing observation of individual ARs in an effort to establish whether any general conclusions can be deduced that relate to the contribution by AR upflows to the slow solar wind.

## Acknowledgements

The work of DHB was performed under contract with the US Naval Research Laboratory and was funded by the NASA *Hinode* programme. The research leading to these results has received funding from the European Commission's Seventh Framework Programme under the grant agreement No. 284461 (eHEROES project). DB's and LvDG's work was supported by the STFC Consolidated Grant ST/H00260X/1. LvDG acknowledges the Hungarian research grant OTKA K-081421. CHM acknowledges financial support from the Argentinian grants UBACyT 20020100100733 (UBA) and PIP 2009-100766 (CONICET). CHM is member of the Carrera del Investigador Científico (CONICET). The work of LZ was supported by NASA grants NNX11AC20G, NNX13AH66G and NNX09AH72G.